# Exchange bias in a mixed metal oxide based magnetocaloric compound YFe$_{0.5}$Cr$_{0.5}$O$_3$

**Mohit K Sharma, Karan Singh and K Mukherjee***

School of Basic Sciences, Indian Institute of Technology Mandi, Mandi-175005, Himachal Pradesh, India

*Email: kaustav@iitmandi.ac.in

## ABSTRACT

We report a detailed investigation of magnetization, magnetocaloric effect and exchange bias studies on a mixed metal oxide YFe$_{0.5}$Cr$_{0.5}$O$_3$ belonging to perovskite family. Our results reveal that the compound is in canted magnetic state (CMS) where ferromagnetic correlations are present in an antiferromagnetic state. Magnetic entropy change of this compound follows a power law ($\Delta S_M \sim H^m$) dependence of magnetic field. In this compound, inverse magnetocaloric effect (IMCE) is observed below 260 K while conventional magnetocaloric effect (CMCE) above it. The exponent '$m$' is found to be independent of temperature and field only in the IMCE region. Investigation of temperature and magnetic field dependence studies of exchange bias, reveal a competition between effective Zeeman energy of the ferromagnetic regions and anisotropic exchange energy at the interface between ferromagnetic and antiferromagnetic regions. Variation of exchange bias due to temperature and field cycling is also investigated.

## 1. Introduction

Mixed metal oxides belonging to perovskite structure have attracted considerable attention in recent years by virtue of their interesting magnetic properties, which can be useful from the viewpoint of technological application and fundamental physics. The combination of 3d-3d or 3d-4d/5d element in a perovskite structured oxide of the form $AB'_{1-x}B''_xO_3$ (A = rare-earth ions, B'/B"= transition metal ions) forms one such mixed metal oxides. Investigations are being carried out on this new kind of half doped chromites and ferrites, $LnFe_{0.5}Cr_{0.5}O_3$, where Ln=La, Y and Dy [1-7]. Combining the two transitions metal within the perovskite structure can be an effective approach of enhance the magnetic property and at the same time tune/induce functional properties as compared to their parent compounds. For example $DyCrO_3$ shows large magnetocaloric effect (MCE) while $DyFeO_3$ shows magnetic field induced multiferroicity below Dy ordering temperature. However, $DyFe_{0.5}Cr_{0.5}O_3$ is significantly different showing a large magnetization and MCE enhanced by magnetoelectric coupling [6]. The compound $YFeO_3$ [8] and $YCrO_3$ [9] exhibit antiferromagnetic ordering around 640 and 140 K respectively along with a weak ferromagnetic behaviour with no functional properties. In contrast, $YFe_{0.5}Cr_{0.5}O_3$ shows the phenomenon of magnetization reversal with a high compensation temperature [2].

Materials exhibiting magnetocaloric effect (MCE) is an active area of research in the area of magnetic condensed matter physics [10], as such materials can be used for solid-state cooling techniques. Such techniques offer a smart solution to the issues related to gas compression/expansion cycle. Additionally, a detailed study of the temperature and field dependence of MCE is helpful in gaining insight about the magnetic phase in a magnetic material and can also provide information about the performance of the material as a magnetic refrigerant.

In magnetic materials, presence of simultaneous ferromagnetic (FM)/antiferromagnetic(AFM) coupling results in the shift of magnetic hysteresis loop away from the centre of symmetry from its normal position when the loop taken in the different cooling field with respect to that taken in zero cooling field [11]. This phenomenon is known as Exchange bias (EB). EB has already been reported in some phase-separated bulk materials, FM/AFM multilayers and magnetic nanoparticles [12-15]. FM/AFM systems are important in understanding the core issues related to exchange bias such as possible origins of the hysteresis loop asymmetry [16-18]. Investigations of EB and its variation with physical parameters have attracted considerable attention in the field of magnetism due to their application in the fundamental science, ultrahigh density magnetic recording, giant magnetoresistance, spin valve, magnetic storage devices, magnetic switches, and magnetic random access memories [2, 19-22]. Systems showing magnetization reversal are also useful for such applications. Recent studies have reported the coexistence of magnetization reversal and exchange bias in some transition metal oxides with the perovskite structure, e.g., $La_{0.2}Ce_{0.8}CrO_3$, $Sr_2YbRuO_6$, and $YbCrO_3$, [23–25]. It would be of great interest to study a compound, where, there is a coexistence of magnetization reversal and EB.

In this context, $YFe_{0.5}Cr_{0.5}O_3$ is an interesting system. Along with magnetization reversal at low applied field this compound shows the phenomenon of magnetic switching [2]. Observation of ferroelectricity and magnetoelectric effect at the magnetic ordering temperature (~260 K) is reported in compound [7]. Other studies suggest a relaxor-like dielectric behaviour around 507 K, attributed to the disordered nature arising from random distribution of $Fe^{3+}$ and $Cr^{3+}$ ions [4]. However, to the best of our knowledge, a detailed and systematic analysis of the low field magnetic study of this compound under different temperature and magnetic field protocol is lacking in literature. Low field magnetic measurements are useful to identify the intrinsic signature of inhomogeneously magnetized system as high magnetic field can mask it. Even though some of the functional properties like observation of magnetocaloric effect [2] and exchange bias [26] is reported in literature, a detailed analysis of these properties are yet to be carried out.

In this work, through bulk magnetization measurements we investigate magnetic state of the sample $YFe_{0.5}Cr_{0.5}O_3$. Our results reveal that the compound is in a canted magnetic state (CMS) where ferromagnetic correlations coexist with antiferromagnetic correlations. Magnetocaloric properties of the compound is also studied which reveals that magnetic entropy change of this compound follows a power law dependence of magnetic field of the form $\Delta S_M \sim H^m$. This compound exhibited inverse magnetocaloric effect (IMCE) below 260 K while conventional magnetocaloric effect above it (CMCE). The exponent $m$ is found to be independent of temperature and field only in the IMCE region. Additionally, temperature and magnetic field dependence studies of exchange bias reveal a competition between effective

Zeeman energy of the FM regions and anisotropic exchange energy between FM/AFM interfaces. Effects of temperature and field cycles on exchange bias field are also investigated

2. **Experimental**

The compound YFe$_{0.5}$Cr$_{0.5}$O$_3$ (YFCO) is prepared by solid state reaction method. Y$_2$O$_3$, Fe$_2$O$_3$ and Cr$_2$O$_3$ procured from Sigma-Aldrich (purity > 99.9%) were taken in stoichiometric amount. After mixing, the powder is subjected to a heat treatment of 900 °C for 24 h. After that powder is regrind and heated at 1200 °C for 24 h. The resulting powder pressed in to pellets and sintered at 1300 °C for 24 h. The x-ray diffraction measurement at room temperature is carried out using Rigaku Smart Lab diffractometer with CuKα radiation (data is taken in angular step 0.02). The Rietveld refinement of the powder diffraction data of YFCO is performed by FullProf Suite software [27]. Temperature and magnetic field dependent magnetization data in the temperature range 2-390 K and magnetic field upto 50 kOe is collected by the Magnetic Property Measurements System (MPMS) from Quantum design, USA. Heat capacity measurement is performed using Physical Property Measurements System (PPMS) from Quantum design in zero magnetic field in temperature range 2-270 K.

3. **Results and discussions**

Fig. 1 shows the representative XRD pattern of YFCO. The sample is seen to be single phase and the XRD pattern is analysed by Rietveld profile refinement [28]. The analysis reveals that YFCO has orthorhombic crystal structure with Pnma space group. The lattice parameters, unit cell volume and atomic positions obtained from Rietveld refinement of powder XRD data is tabulated in Table 1. The obtained lattice parameters match well with the reports on YFCO [2, 4].

Even though there are couple of reports [2, 7] about the temperature response of magnetization of this compound, we repeat this measurement under different temperature and field protocols. In Fig. 2, magnetization plot of YFCO in the temperature range 2 K to 395 K under the Zero Field Cooling (ZFC), Field Cooled Cooling (FCC) and Field Cooled Warming (FCW) condition, at 100 Oe field is shown. In this compound magnetic ordering starts ~ 275 K [obtained from d(*M/H*)/dT vs. T plot (not shown)]. The ordering temperature is in analogy with Ref [2]. It is observed that under ZFC condition, the magnetization decreases as temperature is decreased, goes through a minimum ~ 267 K and then increases with decreasing temperature. At small applied field the minimum occur with a negative magnetization. Bifurcation between ZFC and FCC curve starts around 380 K. FCC curve increases with decreasing temperature and attains a maximum value at ~267 K. Then it decreases and attains a zero value of magnetization at the compensation temperature ($T_{comp}$). Below $T_{comp}$, magnetization is negative down to 2 K. Such behaviour arises due to the fact, that, the net magnetization arising out of coupling between Fe-O-Fe, Cr-O-Cr and Fe-O-Cr is aligned opposite to the applied field. FCC curve obtained under -100 Oe is exactly the mirror image of that obtained under +100 Oe (Lower inset of Fig. 2). The observed behaviour is similar to that reported in Ref [2]. Also from Fig. 2, it is seen that FCW curve follows the FCC curve implying the absence of thermal hysteresis in this compound. $T_{comp}$ decreases with increasing applied field as shown in upper inset of Fig. 2. It is observed that the negative magnetization is suppressed and vanishes for fields greater than ~1.6kOe. It suggests that the net magnetic moments, initially in the opposite direction of the field are realigned in the same direction in a higher magnetic field. The actual $T_{comp}$ = 240 K, is obtained by taking the extrapolated value at $H$ = 0 Oe. Curie-Weiss fit ($M/H = C/(T - \theta_p)$, $C$= constant) of the inverse magnetic susceptibility curve at 5 kOe (not shown) yield a Curie-Weiss temperature $\theta_p$ of ~ -236 K. The negative value of $\theta_p$ indicates dominance of antiferromagnetic interactions. The experimentally obtained value of effective moment is ~ 3.9$\mu_B$ which is less than the value obtained from theoretical calculations (~ 5$\mu_B$) [29]. This indicates that magnetic correlation persist upto the highest measurement temperature. This mismatch between experimental and theoretical moments is not unusual and has been reported in other similar compounds [1, 30].

With a primary aim to understand the magnetic behaviour of the compound more elaborately ZFC isothermal magnetization ($M$) as a function of magnetic field ($H$) is measured at different temperatures. Curves at 10 and 350 K are illustrated in the Fig. 3(a). Virgin $M$-$H$ curves exhibit the linear increment in magnetization with magnetic field without any signature of saturation at all temperatures (Fig. 3(b)). Such linear variation with a non-saturation tendency is generally expected for systems having antiferromagnetic correlations. However, interestingly a magnetic hysteresis is observed at all temperatures up to 350 K. Fig. 3(c) shows the coercive field plotted as a function of temperature. Coercivity decreases with increasing temperature, however still having a significant value at 350 K. Such features support the presence of ferromagnetic correlations among spins in this compound. However, this compound does not show any evidence of metamagnetic transition. Temperature dependence of heat capacity plot, Fig. 3(d), shows absence of any peak. This observation implies the absence of any long range magnetic ordering in this compound. Hence from the above observations it can be said that YFCO compound exhibits magnetic transition from paramagnetic to antiferromagnetic-like canted magnetic state (CMS). The CMS is an inhomogeneous magnetic phase which arises due to the presence of ferromagnetic correlations in an AFM state.

We further investigated the temperature response of the magnetocaloric effect (MCE) of this compound. MCE refer to the change in isothermal magnetic entropy produced by changes in applied magnetic field. When the sample is subjected to a variation of the magnetic field in an isothermal process, the reversible change in entropy $\Delta S$ is equal to the magnetic entropy change $\Delta S_M$, and is calculated using Maxwell's equation:

$$[\partial S/\partial H]_T = [\partial M/\partial T]_H \ldots\ldots\ldots\ldots(1)$$

The temperature response of $-\Delta S_M$ in the field range 0.1- 5 kOe is shown in Fig. 4(a). The nature of MCE is similar to that observed in Ref [2]. The positive value of $\Delta S_M$ below 267 K is a signature of inverse magnetocaloric effect (IMCE) due to antiferromagnetic correlations among spins. With the increase in magnetic field oriental disorder of the magnetic spins aligned opposite to magnetic field increases, resulting in increase of magnetic entropy [31]. Above 260 K, $-\Delta S_M$ is positive, implying a spin alignment. The value increases with applied field as higher field aids to more ordered configuration, resulting in decrease in magnetic entropy. Magnetic field response of $\Delta S_M$ generally follows a power law of the form

$$\Delta S_M \sim H^m \ \ldots\ldots\ldots\ldots (2)$$

Where, $m$ is an exponent depending upon temperature and field [30, 32]. From fig.4 (b), it is observed that $\Delta S_M$ follows the power law given by equation (2) and that the variation of $\ln|\Delta S_M|$ with $\ln H$ is linear. The value of exponent $m$ is ~ 1.1 ± 0.1, for the entire temperature range of measurement below 260 K. For antiferromagnetic systems the value of $m$ is ~ 2. The observed value of *m* is possibly due to the presence of canted magnetic phase in YFCO. To illustrate the magnetic field dependence of *m* for IMCE, *m* is calculated using the formula:

$$m = d \ln |\Delta S_M| / d \ln H \quad \ldots\ldots (3)$$

Field response of *m* is illustrated in inset of Fig. 4(b) and the value of m is found to be ~ $1.1 \pm 0.2$. For both case the fluctuation increases at higher magnetic field. Thus, below 260 K, *m* is found to be temperature and field independent which is a characteristics of inverse magnetocaloric effect (IMCE). Above 260 K, it is found that the value of *m* is dependent of temperature and magnetic field, a characteristic of conventional magnetocaloric effect (CMCE). The compound which exhibits both IMCE and CMCE can be used for a constant temperature bath [33]. Another observation of Fig. 4(a) is that the magnitude of $\Delta S_M$ is quite low. Even in a field of 30 kOe, $\Delta S_M$ is found to be around 0.007 J/kg-K. Generally for a long range ordered magnetic system we have a very high value of $\Delta S_M$. Such low value of $\Delta S_M$ in our case reaffirms that the magnetic state of the sample is in canted magnetic state.

Exchange bias (EB) has been studied extensively in the magnetic system, which have the coexistence of ferro/antiferromagnetism [34]. As YFCO belong to this category of magnetic system, exchange bias expected in this compound. The shifting in the hysteresis loop from the zero field cooling and increase or decrease in the loop width is due to the exchange interaction phenomenon known as the "Exchange bias". This shifting is due to unidirectional anisotropy which arises due to the coupling of FM spin with that of an AFM spin at the interface [35]. To investigate the exchange bias, *M-H* data are taken at *T*~10 K in ZFC condition and after FCC condition. For each FCC *M-H* measurement, sample is warmed in zero field upto 396 K and cooled down to 10 K under the applied field. Fig. 5(a) shows two such *M-H* loop measured under both ZFC and FC ($H = 1$ kOe) conditions. It is observed that FC hysteresis loop shift in magnetization and field axis. This signature highlights the presence of EB phenomenon in this compound. For qualitative comparison of the magnitude of EB effect, cooling field and temperature dependence of EB is carried out. For the former study the sample is cooled from 390 to 10 K under different cooling field ($\leq 4.5$ kOe) and the hysteresis loop is measured between ±5 kOe. The exchange bias field ($H_E$) and remanence exchange bias ($M_E$) is calculated using the formula:

$$H_E = (H_{C1}+H_{C2})/2 \text{ and } M_E = (M_{C1}+M_{C2})/2 \quad \ldots\ldots (4)$$

Where, $H_{C1}$ and $H_{C2}$ are left (-) and right (+) coercive fields respectively $M_{C1}$ and $M_{C2}$ are positive and negative remanence magnetization respectively. Fig. 5(b) shows the $H_E$ and $M_E$ plotted as a function cooling field ($H_{cool}$). It is observed that at low applied FC field values, $H_E$ and $M_E$ increase, but above 1.6 kOe both these parameters decreases. The above observation can be explained in terms of competition between effective Zeeman energy and exchange energy at the interface between ferromagnetic and antiferromagnetic regions. As observed $H_E$ increases with $H_{cool}$, indicating that the interfacial exchange energy is dominating the effective Zeeman energy. This statement is also substantiated from the FCC magnetization verses temperature curve where at low cooling field (<1.6 kOe), the spins are aligned opposite to field direction, resulting in negative magnetization. Hence field below 1.6 kOe is not sufficient to aid the growth of ferromagnetically aligned spins. Due to the dominance of interfacial exchange energy $M_E$ also increases. However, for $H_{cool} \geq 1.6$ kOe, the alignment of FM spins increases. This reduces the influence of the interfacial exchange energy whereas the influence of effective Zeeman energy increases. The spins against the field also align in direction of the cooling field, which influence exchange bias to decrease smoothly. Due to the change in dominating energy $H_E$ and $M_E$ decreases beyond ~1.6 kOe. This fact is substantiated from coercivity ($H_C$) vs. $H_{cool}$ graph (inset of Fig. 5(b)) where $H_C$ increases above ~1.2 kOe, indicating the increase in FM region, resulting in the increase of the effective Zeeman energy resulting in its dominance over anisotropy energy above ~1.6 kOe. This is in analogy to the model proposed for FM/AFM thin films by Meiklejohn and Bean [34]. The exchange bias is inversely proportional to the thickness of FM layer by the relation:

$$H_E = -J S_{AFM} S_{FM}/\mu_0 t_{FM} M_{FM} \quad \ldots\ldots\ldots (5)$$

Where, $J$ is the exchange integral across the interface, $S_{FM}$ and $S_{AFM}$ are interface magnetization of ferromagnet and antiferromagnet respectively, $M_{FM}$ and $t_{FM}$ are magnetization and thickness of FM region. For YFCO, $J< 0$ since there is antiferromagnetic exchange interaction at interface of FM/AFM. At low cooling field $H_E$ increases due to dominance of $S_{AFM}$. But at high cooling field, size of FM region increases ($t_{FM}$) along with $M_{FM}$. Hence, the resulting weak coupling at FM/AFM interface is responsible for decrease the $H_E$ above 1.5 kOe. Niebieskikwiat and Salmon [36] for low cooling fields have indicated a direct reference between $M_E$ and $H_E$ by the relation

$$H_E \sim -M_E/M_S \quad \ldots\ldots\ldots (6)$$

Where, $M_S$ is the saturation magnetization. Our experimental results are in analogy with this relation. Fig. 5(c) shows the results of temperature dependence of the exchange bias. For such measurements, the sample is cooled down from 395 K to measuring temperature under the applied field (4.5 kOe). $H_E$ and $M_E$ is calculated by using equation (4). It is observed that $H_E$ and $M_E$ is temperature dependent and both the parameters increases with as temperature is decreased from 235 to180 K. The observation is due to the dominance of interfacial exchange energy increases with down in temperature. Below 180 K, $H_E$ and $M_E$ decreases as the temperature is decreased as effective Zeeman energy become more effective as compared to interfacial exchange energy. This fact is substantiated from temperature response of $H_C$, where $H_C$ is highest at the lowest temperature. It decreases with rise in temperature resulting in initially dominance of effective Zeeman energy up to 180 K. Above this temperature interfacial exchange energy overcome the dominance of Zeeman energy. The above statements are also supported from the temperature response of $\Delta S_M$ (Fig. 4). Below ~ 235 K, $\Delta S_M$ increases as temperature fall, which indicate more spin misalignment and hence dominance of interfacial energy. As the temperature is decreased below 180 K, $\Delta S_M$ decreases, which indicate more spin ordering, resulting in increase of effective Zeeman energy. Above ~ 235 K, both $H_E$ and $M_E$ changes it sign due to spin realignment as also observed in temperature response of $\Delta S_M$.

In the interface between FM and AFM regions there is unstable state of the spins. Hence a change in $H_E$ is expected when the compound is subjected to temperature and magnetic field cycles (*n*). A gradual change in $H_E$ with increasing *n* is a macroscopic proof rearrangement of spin structure. In order to investigate this in this compound, we take a series of ten *M-H* loops at 10 K using the following experimental protocol. For *n* =1 loop, we cool the sample from 390 K to 10 K and then apply a field of 500 Oe. Then temperature is raised to 390 K. From 390 K the sample is cooled down to 10 K. At 10 K, the *M-H* loop (±50 kOe) is taken. This value of field is chosen below 1.5 kOe, as interfacial exchange interaction is dominant as compared to effective Zeeman energy. Similarly, for *n*= 2, the sample is heated to 390 K in zero field and then cooled to 10 K. At 10 K, 500 Oe field is applied, and then temperature is raised to 390 K and cooled to 10 K. The temperature is again raised to 390 K and cooled to 10 K. At 10 K M-H loop is taken. This protocol is repeated for *n* = 3, 4.............10 [37]. The first and tenth loop is shown in the Fig. 6, highlights a shift in the magnetization and field axis. $H_E$

is seen to decrease as *n* increase. It implies that with increasing number of field and temperature cycling interfacial exchange energy decreases. Consecutive cycling of FM spins results in configuration relaxation of the interfacial AFM spins towards equilibrium and these spins cannot participate in the exchange coupling at the interface. This results in decrease of the exchange bias field with increasing cycling.

## 4. Conclusions

In summary, a detailed investigation of magnetization study on the compound $YFe_{0.5}Cr_{0.5}O_3$, reveal the presence of ferromagnetic correlations along with antiferromagnetic correlations, resulting in canted magnetic state (CMS). In this compound, inverse magnetocaloric effect (IMCE) is observed below 260 K while conventional magnetocaloric effect above it (CMCE). Magnetic entropy change follows the power law dependence of magnetic field and the exponent *m* is found to be independent of temperature and field only in the IMCE region. Temperature and magnetic field response of exchange bias, reveal a competition between effective Zeeman energy of the ferromagnetic regions and exchange energy at the interface between ferromagnetic and antiferromagnetic regions. Also, exchange bias field is seen to reduce as the number of field and temperature cycling increases.

**Acknowledgment**

The authors acknowledge IIT, Mandi for financial support. The experimental facilities of Advanced Material Research Centre (AMRC), IIT Mandi are also being acknowledged.

**References**

[1] A. K. Azad, A. Mellergard, S. G. Eriksson, S. A. Ivanov, S. M. Yunus, F. Lindberg, G. Svensson and R. Mathieu, *Mater. Res. Bull.* 40 (2005) 1633.

[2] J. H. Mao, Y. Sui, X. Q. Zhang, Y. T. Su, X. J. Wang, Z. G. Liu, Y. Wang, R. B. Zhu, Y. Wang, W. F. Liu and J. K. Tang, *Appl. Phys. Lett.* 98 (2011) 192510.

[3] L. H. Yin, W. H. Song, X. L. Jiao, W. B. Wu, L. J. Li, W. Tang, X. B. Zhu, Z. R. Yang, J. M. Dai, L. Zhang and Y. P. Sun, *Solid State Commun.* 150 (2010) 1074-1076.

[4] V. G. Nair, A. Das, V. Subramanian and P. N. Santhosh, *J. Appl. Phys.* 113 (2013) 213907.

[5] V. G. Nair, L. Pal, V. Subramanian and P. N. Santhosh, *J. Appl. Phys.* 115 (2014) 17D728.

[6] L. H. Yin, J. Yang, R. R. Zhang, J. M. Dai, W. H. Song and Y. P. Sun, *Appl. Phys. Lett.* 104 (2014) 032904 and reference therein.

[7] B. Rajeswaran, P. Mandal, Rana Saha, E. Suard, A. Sundaresan and C. N. R. Rao, *Chem Mater.* 24 (2012) 3591.

[8] Y. Ma, X. M. Chen and Y. Q. Lin, *J. Appl. Phys.* 103 (2008) 124111-5.

[9] K. Ramesha, A. Llobet, Th. Proffen, C. R. Serrao and C. N. R. Rao, *J. Phys.: Condens. Matter.* 19 (2007) 102202.

[10] K. A. Gschneidner Jr, V. K. Pecharsky, and A. O. Tsokol, *Rep. Prog. Phys.* 68 (2005) 1479-1539.

[11] W. H. Meiklejohn and C. P. Bean, *Phys. Rev.* 102 (1956) 1413.

[12] J. Nogues and Ivan K. Schuller, *J. Magn. Magn. Mater.* 192 (1999) 203.

[13] B. Martínez, X. Obradors, L. Balcells, A. Rouanet and C. Monty, *Phys. Rev. Lett.* 80 (1998) 181.

[14] Y. K. Tang, Y. Sun and Z. H. Cheng, *Phys. Rev. B* 73 (2006) 174419.

[15] M. Patra, S. Majumdarand, S. Giri, *Solid State Commun*. 149 (2009) 501.

[16] O. Hovorka, A. Berger and G. Friedman, *Appl. Phys. Lett.* 89 (2006) 142513.

[17] O. Hovorka, A. Berger and G. Friedman, *J. Appl. Phys.* 101 (2007) 09E515.

[18] M. Zhu, M. J. Wilson, B. L. Sheu, P. Mitra, P. Schiffer and N. Samarth, *Appl. Phys. Lett.* 91 (2007) 192503.

[19] R. L. Stamps, *J. Phys. D* 33 (2000) R247.

[20] J. C. S. Kools, *IEEE Trans. Magn.* 32 (1996) 3165-3184.

[21] I. L. Prejbeanu, M. Kerekes, R. C. Sousa, H. Sibuet, O. Redon, B. Dieny and J. P. Nozieres, *J. Phys.:Condens. Matter.* 19 (2007) 165218.

[22] P. Mandal, A. Sundaresan, C. N. R. Rao, A. Iyo, P. M. Shirage, Y. Tanaka, Ch. Simon, V. Pralong, O. I. Lebedev, V. Caignaert and B. Raveau, *Phys. Rev. B* 82 (2010) 100416.

[23] P. K. Manna, S. M. Yusuf, R. Shukla and A. K. Tyagi, *Appl. Phys. Lett*. 96 (2010) 242508.

[24] R. P. Singh, C. V. Tomy and A. K. Grover, *Appl. Phys. Lett*. 97 (2010) 182505.

[25] Y. L. Su, J. C. Zhang, Z. J. Feng, L. Li, B. Z. Li, Y. Zhou, Z. P. Chen and S. X. Cao, *J. Appl. Phys.*108 (2010) 013905.

[26] J. H. Mao, Y. Sui, X. Q. Zhang, X. J. Wang, Y. T. Su, Z. G. Liu, Y. Wang, R. B. Zhu, Y. Wang, W. F. Liu, X. Liu, *Solid State Commun.* 151 (2011) 1982-1985.

[27] J. Rodriguez-Carvajal, Abstracts of the Satellite Meeting on Powder Diffraction of the XV Congress of the IUCr (1990) p. 127

[28] R. A. Young, (ed) *The Rietveld Method* (Oxford: Oxford University Press) (1993).

[29] P. K. Vijayanandhini, Ch. Simon, V. Pralong, Y. Bréard, V. Caignaert, B. Raveau, P. Mandal, A. Sundaresan and C. N. R. Rao, *J. Phys.:Condens. Matter.* 21 (2009) 486002.

[30] S. Chandra, A. Biswas, S. Datta, B. Ghosh, A. K. Raychaudhuri and H. Srikanth, *Nanotechnology* 24 (2013) 505712.

[31] P. J von Ranke, N. A. de Oliveira, B. P. Alho, E. J. R. Plaza, V. S. R. de Sousa, L. Caronand M. S. Reis, *J. Phys.: Condens. Matter.* 21 (2009) 056004 (8pp).

[32] A. Biswas, S. Chandra, T. Samanta, B. Ghosh, S. Datta, M. H. Phan, A. K. Raychaudhuri, I. Das and H. Srikanth, *Phys. Rev B* 87 (2013) 134420.

[33] S. M. Yusuf, A. Kumar and J. V. Yakhmi, *Appl. Phys. Lett.* 95 (2009) 182506.

[34] W. H. Meiklejohn and C. P. Bean, *Phys. Rev. Lett.* 105 (1957) 904.

[35] Y. Tang, Y. Sun and Z. Cheng, *Phys. Rev. B* 73 (2006) 174419.

[36] D. Niebieskikwiat and M. B. Salamon, *Phys. Rev. B*72 (2005) 174422.

[37] D. Zhu, V. Hardy, A. Maignan and B. Raveau, *J. Phys.: Condens. Matter.* 16 (2004) L101-L107.

**Table 1:** Unit cell parameter and atomic positional parameter for $YFe_{0.5}Cr_{0.5}O_3$.

| Lattice parameters | | | |
|---|---|---|---|
| $a$ (Å) | $b$ (Å) | $c$ (Å) | $V$ (Å$^3$) |
| 5.5574(3) | 7.5689(7) | 5.2628 (2) | 221.380 (9) |
| Bragg R-factor = 2.93 | $R_f$- factor = 2.54 | | $\chi^2$ = 2.11 |
| | | | Atomic positions | | |
| | Site | x | y | z |
| Y | 4c | 0.0666 | 0.2500 | 0.9832 |
| Fe | 4b | 0.5000 | 0.0000 | 0.0000 |
| Cr | 4b | 0.5000 | 0.0000 | 0.0000 |
| O1 | 4c | 0.4638 | 0.2500 | 0.1071 |
| O2 | 8d | 0.3040 | 0.0551 | 0.6924 |

**Figures:**

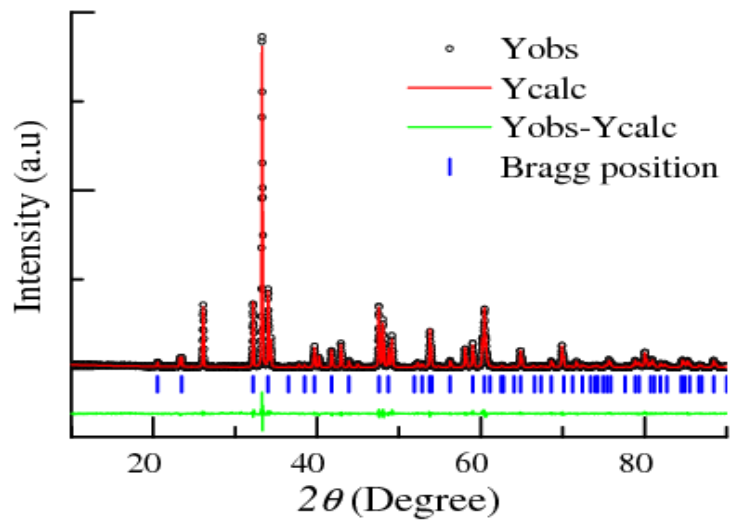

Fig. 1. X-ray diffraction patterns (Cu $K_\alpha$) for $YFe_{0.5}Cr_{0.5}O_3$. The result of Rietveld analysis of the XRD pattern is also shown.

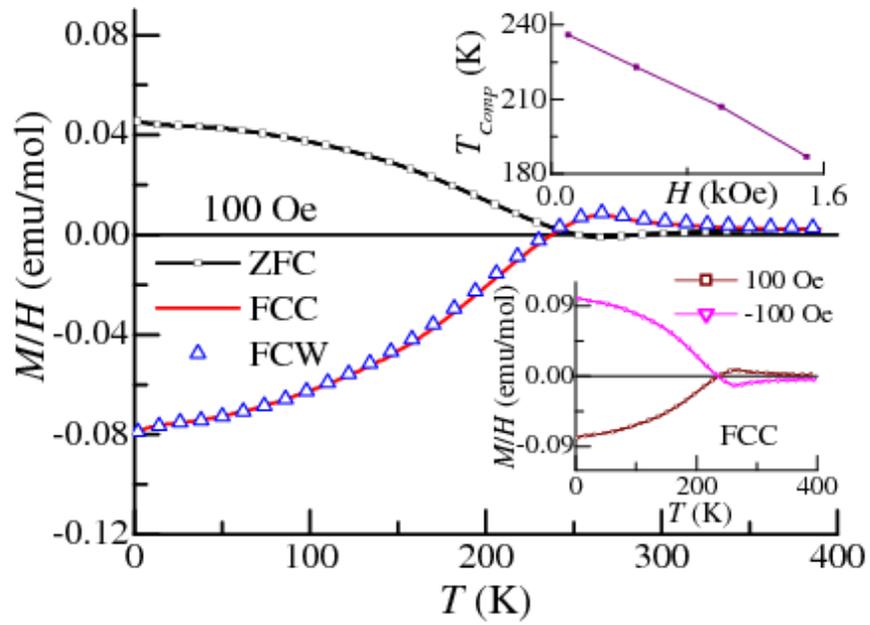

Fig. 2. Temperature dependent magnetization curves (ZFC, FCC and FCW) at 100 Oe for YFe$_{0.5}$Cr$_{0.5}$O$_3$. Lower inset: Temperature response of FCC magnetization at -100 and 100 Oe. Upper inset: Compensation temperature as a function of magnetic field.

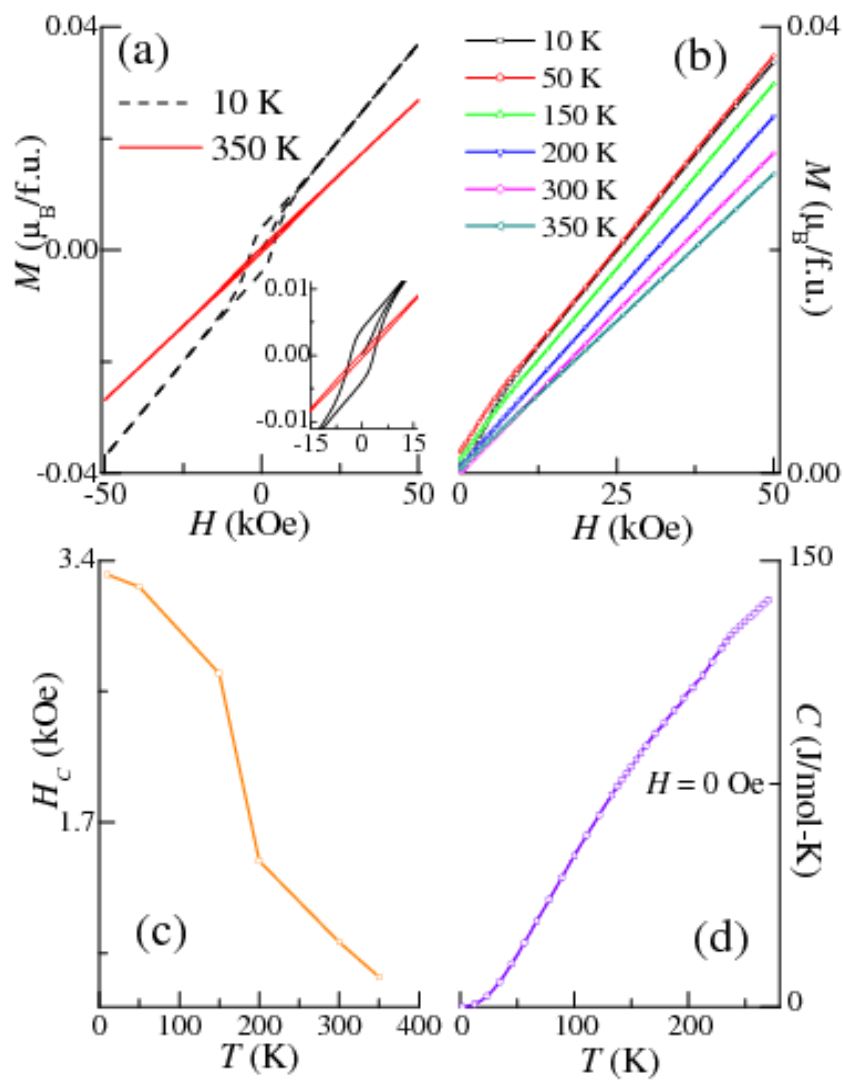

Fig. 3. (a) Magnetic hysteresis behaviour at 10 K and 350 K for field cycling of ±50 kOe for YFe$_{0.5}$Cr$_{0.5}$O$_3$. Inset: Magnified hysteresis loop. (b) Virgin magnetization curve as a function of magnetic field at different temperature. (c) Temperature dependence of coercive field. (d) Heat capacity plotted as a function of temperature in zero magnetic field. Unless stated, the lines through the data points serve as guides to the eyes.

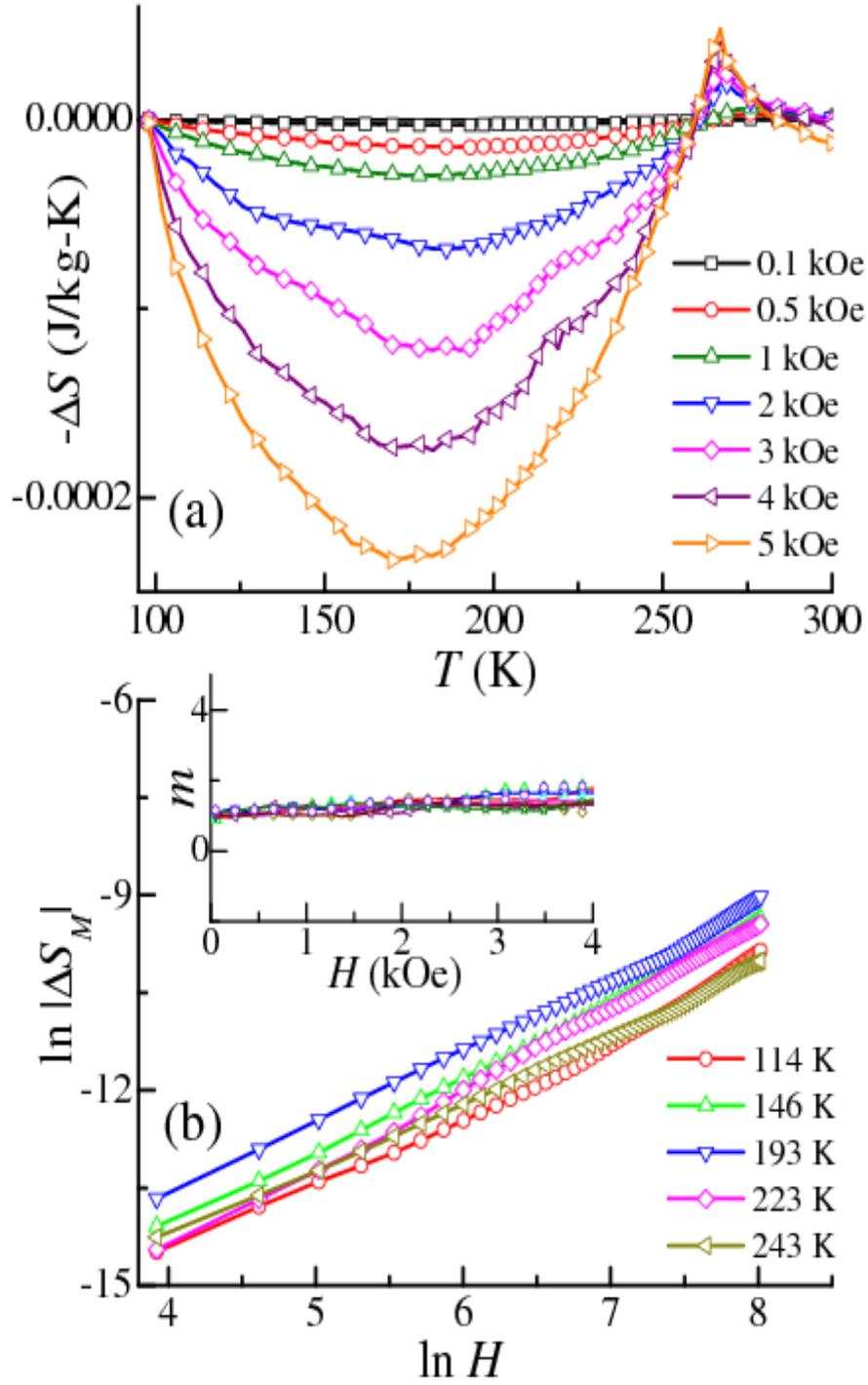

Fig. 4. For YFe$_{0.5}$Cr$_{0.5}$O$_3$, (a) Isothermal magnetic entropy change (-$\Delta S_M$) for different magnetic field (0.1 to 5 kOe) plotted as a function of temperature. (b) ln|$\Delta S_M$| vs. ln|$H$| curves in the range ~ 0 to 5 kOe at different temperatures. Inset: Magnetic field dependence of $m$ [in accordance to equation (3)] at different temperatures (110-250 K).

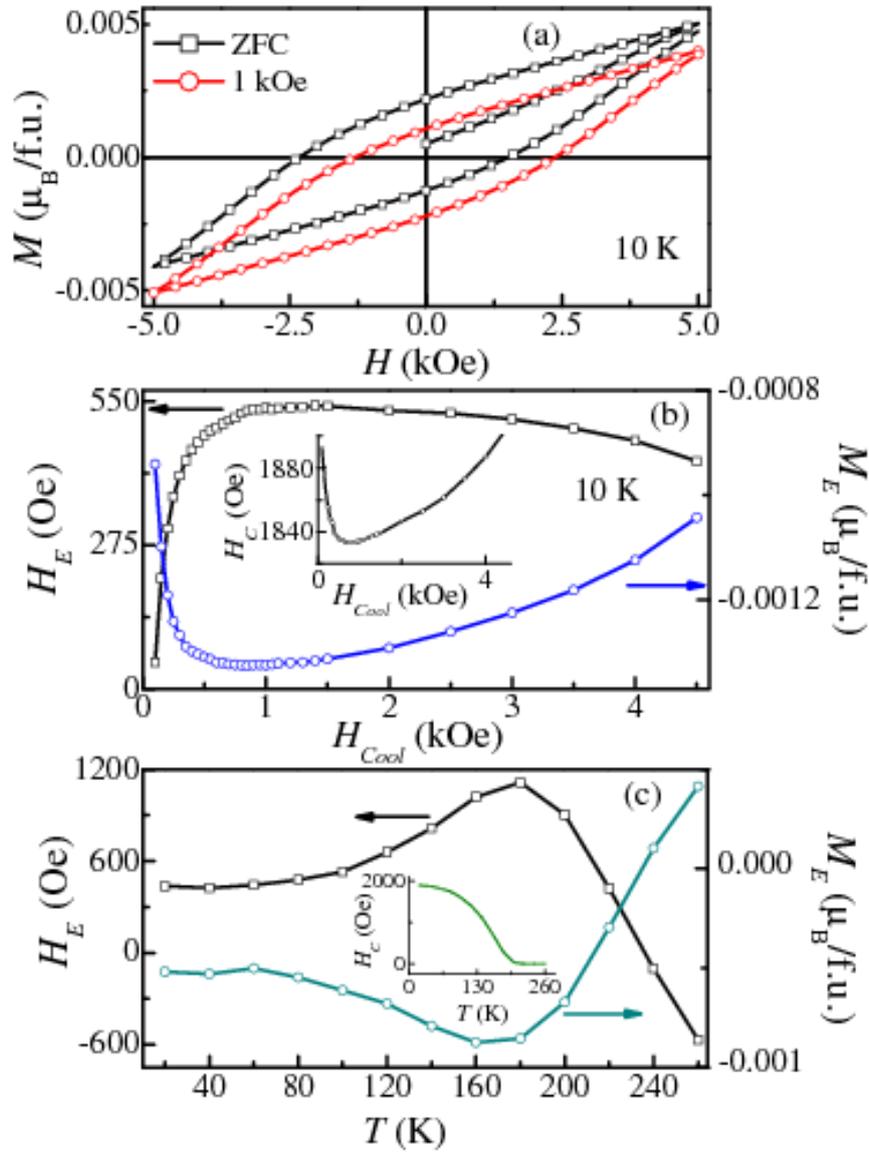

Fig. 5. For YFe$_{0.5}$Cr$_{0.5}$O$_3$, (a)Magnetic hysteresis loops for zero field cooling and field cooling in 1 kOe at 10 K. (b) Exchange bias plotted as a function of cooling field at 10 K. Inset: Coercive field plotted at a function of cooling field at 10K. (c) Temperature response of exchange bias after field cooling at 4.5kOe. Inset: Coercive field plotted as a function of temperature after field cooling at 4.5 kOe.

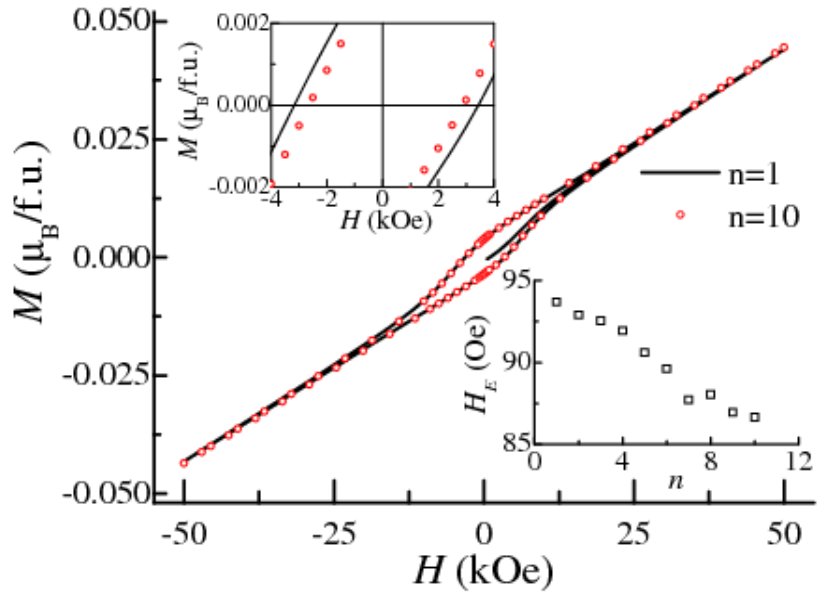

Fig. 6. For YFe$_{0.5}$Cr$_{0.5}$O$_3$, Hysteresis loops for the first and the tenth loop at 10 K. Upper inset: hysteresis loop figure shows the shifting between first and the tenth loop. Lower inset: Exchange bias plotted as a function of *n* at 10 K.